\documentclass{aa}  

\usepackage{graphicx}
\usepackage{txfonts}
\usepackage{natbib}
\bibpunct{(}{)}{;}{a}{}{,}
\begin{document} 

   \title{Thermal-nonthermal energy partition in solar flares derived from X-ray, EUV, and bolometric observations}
   \subtitle{Discussion of recent studies}
   
     \author{A. Warmuth \inst{1} \and G. Mann\inst{1}
   }
   \offprints{A. Warmuth, \email{awarmuth@aip.de}}

   \institute{Leibniz-Institut f\"ur Astrophysik Potsdam (AIP), An der Sternwarte 16, 14482 Potsdam, Germany}

   \date{Received <date> / Accepted <date>}

   \authorrunning{A. Warmuth\and G. Mann}
   \titlerunning{Thermal-nonthermal energy partition in solar flares}
 
  \abstract
   {In solar flares, energy is released impulsively and is partly converted into thermal energy of hot plasmas and kinetic energy of accelerated nonthermal particles. It is crucial to constrain the partition of these two energy components to understand energy release and transport as well as particle acceleration in solar flares. Despite numerous efforts, no consensus on quantifying this energy balance has yet been reached.} 
   {We aim to understand the reasons for the contradicting results on energy partition obtained by various recent studies. The overarching question we address is whether there is sufficient energy in nonthermal particles to account for the thermal flare component.}
   {We considered five recent studies that address the thermal-nonthermal energy partition in solar flares. Their results are reviewed, and their methods are compared and discussed in detail.}
   {The main uncertainties in deriving the energy partition are identified as (a) the derivation of the differential emission measure (DEM) distribution and (b) the role of the conductive energy loss for the thermal component, as well as (c) the determination of the low-energy cutoff for the injected electrons. The bolometric radiated energy, as a proxy for the total energy released in the flare, is a useful independent constraint on both thermal and nonthermal energetics. In most of the cases, the derived energetics are consistent with this constraint. There are indications that the thermal-nonthermal energy partition changes with flare strength: in weak flares, there appears to be a deficit of energetic electrons, while the injected nonthermal energy is sufficient to account for the thermal component in strong flares. This behavior is identified as the main cause of the dissimilar results in the studies we considered. The changing partition has two important consequences: (a) an additional direct (i.e. non-beam) heating mechanism has to be present, and (b) considering that the bolometric emission originates mainly from deeper atmospheric layers, conduction or waves are required as additional energy transport mechanisms.}
   {}

    \keywords{Sun: flares -- Sun: X-rays, gamma rays -- acceleration of particles}

\maketitle

%

\section{Introduction}
\label{sec:intro}

In solar eruptive events such as flares \citep[e.g.][]{Fletcher11} and coronal mass ejections \citep[CMEs; e.g.,][]{Chen11b,Webb12}, a large amount of energy ($\leq$ 10$^{33}$\,erg) that is originally stored in nonpotential coronal magnetic fields is released impulsively and converted into other forms of energy,  presumably triggered by magnetic reconnection \citep[e.g.,][]{Priest82}. Newly reconnected magnetic field lines rapidly move away from the reconnection site, taking the plasma with them. This forms two outflow jets. The outflow may also be heated, for example, by standing slow-mode shocks that separate the inflow from the outflow in Petschek-style reconnection \citep[cf.][]{Cargill82}, as well as by fast-mode termination shocks. Somewhere in this geometry, efficient particle acceleration due to an as yet poorly understood mechanism \citep[cf.][]{Zharkova11,Mann15}, is taking place. The downward-moving field lines form flaring loops that become filled with dense plasma that is evaporated from the chromosphere by strong heating. In major eruptive events, the unstable magnetic structure forms a flux rope (in which a filament may be embedded) that is expelled from the corona. The upward-moving reconnected field lines become part of the flux rope, which is subsequently observed as a CME. This general scenario is supported by many observations and represents our standard model of a solar eruption. In its 2D form, it is known as the CSHKP model \citep[cf.][]{Carmichael64, Sturrock66, Hirayama74, Kopp76}. Recently, the standard model has been extended to 3D \citep[see][]{Aulanier12,Aulanier13,Janvier13}.

It is quite evident that a solar eruptive event is characterized by a complex scenario of energy release, transport, and conversion. This starts with the free magnetic energy in the flaring active region and continues with the amount of energy that is actually released, the kinetic and thermal energy of the reconnection outflow jets, the kinetic energy in accelerated particles and in evaporation flows, the thermal energy of evaporated plasma, and radiative and conductive energy losses of various plasmas. In the case of eruptive flares, the kinetic and potential energy of CMEs, the energy of CME-driven shocks, and finally, the energy content in solar energetic particles (SEPs) is added to this. 

A quantitative characterization of the different forms of energy therefore represents a crucial observational constraint for models of solar eruptions in general, as well as for magnetic reconnection, heating, and particle-acceleration processes in particular. Several studies have tried to characterize the partition between subsets of these energies in solar flares or eruptive events. In this context, three questions have attracted particular interest: (i) whether there is enough free magnetic energy in an active region to account for the total energy released in a flare or CME, (ii) what the energy partition is between flare and CME, and (iii) whether nonthermal particles can power the thermal component in flares. It is generally found that enough free magnetic energy is available to drive flares and CMEs \citep{Emslie12,Aschwanden17a}. With regard to the partition between energy of the flare and the associated CME, the situation is less clear. \cite{Emslie12} found energies on the same order of magnitude, and \cite{Aschwanden17a} concluded that the flare component dominates the energetics.

\begin{table*}[ttt!!!]
\caption[]{Overview of the flare energetics studies discussed. For details, see the main text.}
         \centering
\begin{tabular}{lcccccc}
        \hline
        \hline
            \noalign{\smallskip}
 & no. & GOES class & thermal & data source & data source & thermal  \\
study & flares & range & model & thermal spectrum & thermal volume & losses \\
        \noalign{\smallskip}
            \hline
            \noalign{\smallskip}
Stoiser et al. 2007 (S+07) & 18 & A3--B7 & isoth. & RHESSI & TRACE & -- \\
Emslie et al. 2012 (E+12) & 38 & C5--X28 & isotherm. & RHESSI & RHESSI & rad.  \\
Inglis \& Christe 2014 (IC14) & 10 & B3--B9 & multitherm. & RHESSI+AIA & RHESSI & rad.  \\
Warmuth \& Mann 2016 (WM16) & 24 & C3--X17 & isotherm. & RHESSI+GOES & RHESSI & rad., cond. \\
Aschwanden et al. 2017 (A+17) & 188 & M1--X7 & multitherm. & AIA & AIA & rad.  \\
        \noalign{\smallskip}
            \hline
\end{tabular}
\label{tab:studies}
\end{table*}

In this study, we focus on the last question, whether there is enough energy in nonthermal particles to heat the thermal plasma that is observed in solar flares. Hard X-ray (HXR) and gamma-ray observations clearly demonstrate that electrons and ions are efficiently accelerated to high energies during solar flares \citep[cf.][]{Holman11,Vilmer11}. The most widely accepted mechanism for the generation of the thermal flare plasma is chromospheric evaporation by electron beams. This scenario is supported by the Neupert effect \citep{Neupert68}, which refers to the observation that the time profile of nonthermal HXR or microwave emission tends to closely match the time derivative of the (thermal) soft X-ray (SXR) flux. This implies that the energy is first released in the form of nonthermal electrons, which then follow the reconnected magnetic field lines down to denser layers of the atmosphere where they thermalize and initiate chromospheric evaporation that then fills up the flaring loops with SXR-emitting plasma. The electron beam scenario is further supported by spatial and temporal correlations of the nonthermal HXR emission and evaporation signatures such as hot upflows \citep[e.g.,][]{Milligan06a,Tian14,LiDong15}.

There is no consensus so far about the answer to the simple question whether nonthermal electrons can provide sufficient energy to power the thermal flare component because even the most recent studies give contradicting results. In Sect.~\ref{sec:studies} we briefly describe five relevant studies, their method, and the main conclusions. In Sect.~\ref{sec:disc} we assess the treatment of various specific issues in the different studies, which allows us to set the conflicting results into perspective and derive a possible way forward. The conclusions are given in Sect.~\ref{sec:conc}


\section{Overview of studies}
\label{sec:studies}

In the following, we describe five studies from the past 15 years that meet two criteria: (a) they simultaneously determined the thermal energy of the hot plasma and the energy input by nonthermal electrons in the same flare events, and (b) they considered a larger sample of flares in order to obtain statistically valid results and investigate correlations between different flare parameters. While most of the selected studies have addressed several different topics and several of them have derived additional flare and/or CME energetics, we focus on the issue of the partition between thermal and nonthermal flare energy here.

As an overview, Table~\ref{tab:studies} shows some basic characteristics of these studies. This includes the number of events studied, the range of SXR flare importance of the selected flares as measured by the Geostationary Orbiting Environmental Satellites (GOES), the model used to characterize the thermal component (isothermal vs. multithermal), the data source for the spectral and geometric parameters of the thermal component, and the energy-loss processes of the thermal plasma that were considered (radiation and conduction).

\subsection{Stoiser et al. 2007 (S+07)}
\label{sec:stoiser07}

\citet[][henceforth referred to as S+07]{Stoiser07} have derived thermal and nonthermal energies for 18 microflares (with background-subtracted GOES classes ranging from A3 to B7) that occurred within a single active region on {\mbox 2003 September 26}. The thermal parameters were derived from isothermal fits to  Reuven Ramaty High Energy Solar Spectroscopic Imager \citep[RHESSI;][]{Lin02} HXR spectra at the flare peaks. The corresponding source volume was estimated from the footpoint brightenings observed at 1600~\AA\ by the Transition Region and Coronal Explorer \citep[TRACE;][]{Handy99} under the assumption of a semicircular loop. The nonthermal energy in electrons was calculated by a power-law fit to the nonthermal part of the photon spectrum, conversion to electron beam power by assuming thick-target emission and a fixed low-energy cutoff of 10~keV, and integration over the time of detectable emission above 10~keV under the assumption of a triangular time profile.
S+07 reported that nonthermal dominates thermal energy, with a median ratio of $\approx$24.

\subsection{Emslie et al. 2012 (E+12)}
\label{sec:emslie12}

The study of \citet[][henceforth E+12]{Emslie12} focused on deriving a broad range of flare, CME, and magnetic field energetics for a sample of 38 larger flares (ranging from C6 to X28 in GOES class). This was achieved by applying the method previously used on two large flares \citep{Emslie04,Emslie05}. The peak thermal energy was deduced from isothermal fits to RHESSI spectra in combination with source areas (and thus volumes) obtained from RHESSI imaging. In addition, the radiative loss of the hot plasma was derived from GOES observations by applying an isothermal fit to the GOES fluxes and using the resulting temperature and emission measure to compute the radiative loss as given by the radiative loss function provided by the CHIANTI atomic database \citep{Dere97}.

The energy in nonthermal electrons was derived from thick-target fits to the RHESSI spectra and integration in time over the whole event. In this process, the highest low-energy cutoffs that were consistent with the data were used. Therefore the derived nonthermal energies are actually a lower estimate.
With respect to the thermal-nonthermal energy partition, the study concluded that the energy in injected electrons is sufficient to heat the hot flare plasma and to account for its radiative losses. 

\subsection{Inglis \& Christe 2014 (IC14)}
\label{sec:inglis14}

The energetics of ten microflares was studied by \citet[][henceforth IC14]{Inglis14}. Thermal energies were calculated from a multithermal model constrained by both RHESSI HXR spectra and extreme ultraviolet (EUV) fluxes from the Solar Dynamics
Observatory \citep[SDO;][]{Pesnell12} Atmospheric Imaging Assembly \citep[AIA;][]{Lemen12}. The thermal plasma volume was obtained from RHESSI thermal source areas. The radiative energy loss was derived from the differential emission measure (DEM) profiles using CHIANTI. The energy in nonthermal electrons was constrained by fitting the high-energy HXR emission not accounted for by the multithermal model with a thick-target bremsstrahlung model (effectively giving an upper estimate for the low-energy cutoff and a lower estimate for the nonthermal power).
IC14 concluded that the minimum nonthermal energy content averages approximately 30\% of the thermal energy deduced from the multithermal model.

\subsection{Warmuth \& Mann 2016a/b (WM16)}
\label{sec:warmuth16}

In a series of studies, Warmuth \& Mann performed a detailed characterization of the geometric \citep{Warmuth13a,Warmuth13b} and spectral parameters \citep{Warmuth16a} of 24 solar flares ranging from small C-class to large X-class flares using HXR imaging and spectroscopy from RHESSI. Based on these parameters, energy partition was studied by \citet[][henceforth WM16]{Warmuth16b}. Thermal energy as a function of time was derived from isothermal fits to the RHESSI spectra combined with RHESSI thermal source sizes. This was supplemented by isothermal fits to the GOES fluxes. Radiative losses were computed from the isothermal parameters provided by RHESSI and GOES using CHIANTI. The thermal parameters were also used to derive the conductive energy loss, assuming Spitzer conductivity with the appropriate saturation limits \citep[cf.][and references therein]{Battaglia09}. Finally, the total heating requirement was obtained from the various thermal energetics. Similar to E+12 and IC14, a lower limit to the energy input by nonthermal electrons was derived from thick-target fits to the RHESSI spectra.

WM16 found that conductive losses are energetically very important. The total heating requirements can only be fulfilled by energetic electrons in stronger flares. In weak flares, the thermal requirements are higher by up to an order of magnitude than the nonthermal input.

\subsection{Aschwanden et al. 2017 (A+17)}
\label{sec:asch17}

In an extensive series of studies on energetics, Aschwanden et al. (2017) quantified the free magnetic energy in active regions \citep{Aschwanden14}, the energy of CMEs \citep{Aschwanden16a}, and the thermal energy of the plasma \citep{Aschwanden15} as well as the energy in nonthermal electrons \citep{Aschwanden16b} in solar flares. The partition between these energies in solar eruptive events is discussed in \cite{Aschwanden17a}. In the following, we collectively refer to the results of these studies as A+17.

The peak thermal energies of 391 M- and X-class flares were computed from DEM profiles obtained with the spatial synthesis DEM method \citep{Aschwanden13a} from AIA images. The flare volumes were derived from the areas in emission measure (EM) maps above a certain threshold. The radiative loss was given by the GOES SXR fluxes under the isothermal assumption using the CHIANTI loss rates. In 191 of these M- and X-class flares, the nonthermal electron energy was derived from collisional thick-target fits to RHESSI spectra, with a low-energy cutoff that was obtained from an analytical approximation to the warm-target model introduced by \citet{Kontar15}. In this study, we only consider the 188 flares for which both thermal and nonthermal energies are available. 

A+17 concluded that the  energy in nonthermal electrons is generally about an order of magnitude higher than the peak thermal energy. Recently, \citet[][henceforth A+19]{Aschwanden19} have constrained the low-energy cutoff and nonthermal electron energetics for the same event sample with four different methods, which has led to some modifications of the original conclusions that we discuss in Sect.~\ref{sec:nth}. 

\subsection{Summary of conclusions on energy partition}
\label{sec:studies_concl}

We summarize the conclusions on thermal-nonthermal energy partition given by the five studies. E+12 find that nonthermal electrons \emph{\textup{can}} account for the thermal plasma, and both S+07 and A+17 conclude that the electrons actually dominate flare energetics. In contrast, IC14 find that there is not enough energy in the nonthermal electrons to power the thermal plasma. Finally, WM16 conclude that the electrons \emph{\textup{can}} account for the thermal plasma \emph{only} in stronger flares. This clearly shows that no consensus on the thermal-nonthermal energy partition in solar flares has been reached yet.


\section{Discussion}
\label{sec:disc}

\subsection{Bolometric radiated energy as an independent constraint}
\label{sec:bol}

The thermal and nonthermal energetics discussed were derived from X-ray and/or EUV observations, while it is well known that solar flares emit copiously at longer wavelengths. A meaningful discussion of energy partition in flares is thus only possible when the total energy released in a solar flare is constrained first.

Based on total solar irradiance (TSI) observations obtained from the Total Irradiance Monitor (TIM) on the Solar Radiation and Climate Experiment (SORCE) spacecraft, the total (bolometric) energy radiated by flares has been measured individually  for a few large X-class flares \citep[][E+12]{Woods06}. Because TSI fluctuations due to solar p-mode oscillations do not allow such measurements in smaller flares, \citet[][]{Kretzschmar10} have applied a superposed epoch analysis to TSI data provided by the Variability of Solar Irradiance and Gravity Oscillations (VIRGO) instrument aboard the  Solar and Heliospheric Observatory (SOHO) spacecraft. In this method, TSI light curves of many flares are overlaid so that they are all centered on the flare peak time. Averaging over the light curves then suppresses the random TSI fluctuations and an average bolometric energy can  be extracted for the flare ensemble. Using four ensembles of varying mean GOES class, \citet{Kretzschmar11} was able to derive bolometric energies from X-class down to C-class flares. When they are plotted as a function of mean GOES peak flux, the bolometric energies of the four event ensembles are well fit by a power law with a slope of $\alpha = 0.79 \pm 0.11,$ which is also consistent with the individual bolometric energies derived from SORCE/TIM. The agreement between the two different instruments and analysis techniques gives us some confidence in the validity of the derived energies.

The bolometric energy is a proxy for the total energy released in solar flares. Regardless of the way in which energy is released in a solar flare (whether by direct heating of plasma, fast bulk flows, or nonthermal particles), in the end, everything is thermalized and radiated away. Because the bolometric energy covers the whole spectrum, we expect it to correspond to the total energy that has been released originally. This refers only to the energy released in the flare and does not account for the energy of an associated CME or filament eruption.

In the following, we therefore frequently compare the results of the five studies to the bolometric energies in order to (i) have an independent consistency check and (ii) assess the fraction of the various energies with respect to the total energy released.

An additional important aspect is that the bolometric emission is dominated by near-UV, white-light, and near-IR radiation that originates from comparatively cool and dense plasmas in the chromosphere and photosphere \citep{Woods06,Kretzschmar11}. Assuming that the primary energy release takes place in the corona, this places some stringent requirements on the processes that transport the energy down to lower atmospheric layers (electrons, ions, conduction, and waves).

\subsection{Event selection issues}
\label{sec:selec}

One potential reason for discrepancies between the various studies are obviously  different event selection criteria. In particular, the samples used differ significantly in terms of flare importance, for instance, as measured by the GOES peak flux. The minimum and maximum GOES class of the various flare samples are given in Table~\ref{tab:studies}. While E+12 and A+17 mostly considered M- and X-class flares and IC14 restricted their study to microflares, WM16 covered the range from C- to X-class flares.

If thermal and nonthermal energetics scale differently with flare importance, then it is natural that different partitions are found for these strongly dissimilar event samples. We address this issue in Sect.~\ref{sec:th_nth} after considering the dependence of thermal and nonthermal energetics on flare class in Sects.~\ref{sec:th} and \ref{sec:nth}.

\subsection{Thermal issues}
\label{sec:th}

\subsubsection{Scaling of thermal energy with GOES class}
\label{sec:th_scaling}

\begin{figure}
\resizebox{\hsize}{!}{\includegraphics{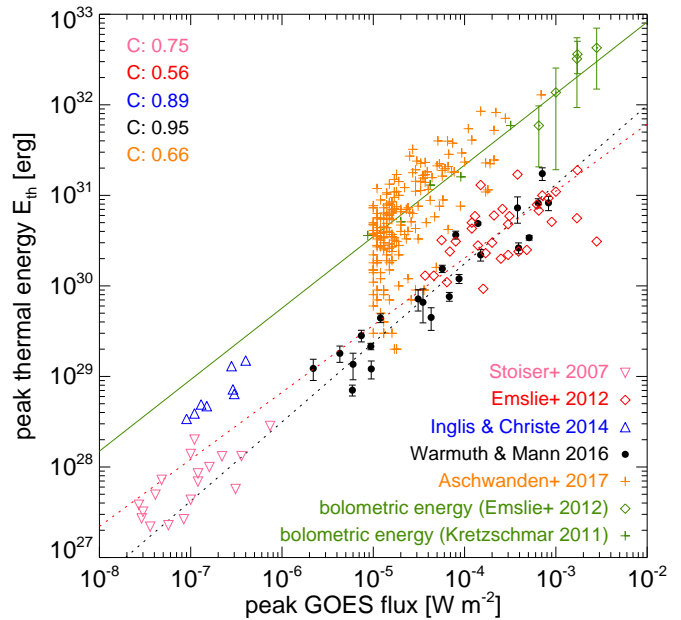}}
\caption{Peak thermal energy $E_\mathrm{th}$ vs. peak GOES flux for all five studies. For comparison, the total radiated energies $E_\mathrm{bol}$ derived from SORCE/TIM (green diamonds) and SOHO/VIRGO (green crosses; the green line is a power-law fit to these data points) are shown. $C$ indicates the linear correlation coefficient of the logarithms of the value pairs. The dotted lines indicate power-law fits to the values of E+12 and WM16.}
\label{fig:eth_goes}
\end{figure}

For an overview of the thermal energetics, we plot the peak thermal energies $E_\mathrm{th}$ as a function of the peak GOES SXR flux as determined by the five studies in Fig.~\ref{fig:eth_goes}. We used the background-subtracted GOES peak fluxes for all studies except for E+12 and A+17, for which only the unsubtracted fluxes are available (the same approach is used throughout this paper). For weaker events, the subtracted fluxes are more meaningful because the background can be quite significant. However, the effect of background subtraction is negligible for the M- and X-class flares that constitute the event samples used by E+12 and A+17, so that we do not introduce a bias by using the raw values for these two studies. 

All studies show a good to excellent correlation with GOES class. E+12 and WM16 are very consistent and are about an order of magnitude lower than the bolometric energies, which are plotted in green for comparison. IC14 appears to be elevated by about half an order of magnitude as compared to extrapolations of E+12 and WM16 to weaker flares (indicated by dotted lines in Fig.~\ref{fig:eth_goes}), while S+07 is very consistent with the extrapolations. Finally, the thermal energies of A+17 are about an order of magnitude higher than those of the other studies, and they are roughly equal to the bolometric energies. The question now is how these differences can be explained.

\subsubsection{Spectral range and thermal model}
\label{sec:th_method}

\begin{figure*}
\centering
\includegraphics[width=17cm]{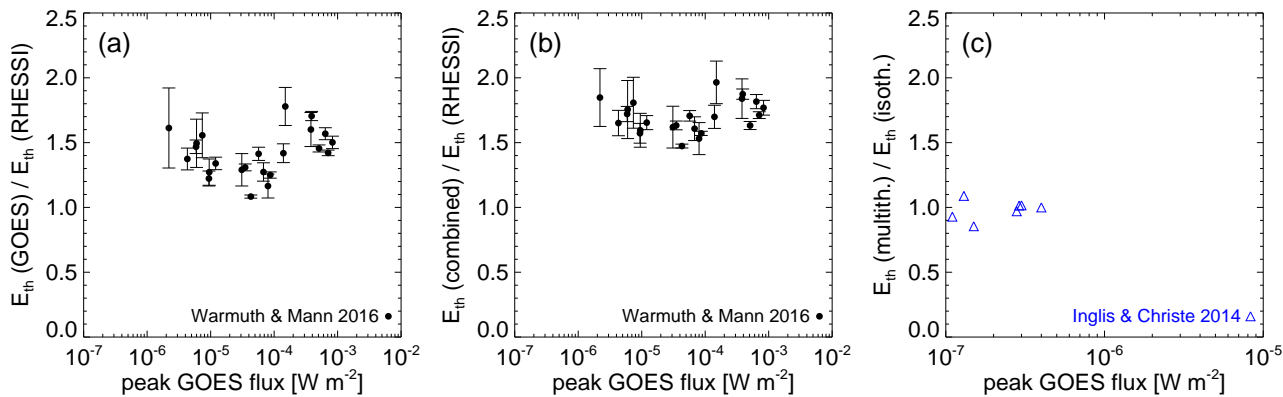}
\caption{Ratios of peak thermal energies $E_\mathrm{th}$ derived with different methods for the same flare samples plotted vs. peak GOES flux. (a) Ratio of GOES- and RHESSI-derived energy in WM16. (b) Ratio of combined (using RHESSI and GOES) and RHESSI-derived energy in WM16. (c) Ratio of multithermal and isothermal energy in IC14.}
\label{fig:eth_hxsx}
\end{figure*}

The five studies used quite diverse approaches to determine the thermal energy of the flare plasma from observations and also used different data analyses, which may lead to systematic differences that might account for the different results on energy partition. 

E+12 used isothermal fits of the RHESSI HXR spectra to obtain emission measure and temperature as input for the derivation of thermal energies. WM16 combined this method with isothermal fits of the GOES fluxes and derived three thermal energies: isothermal GOES, isothermal RHESSI, and a combined value that was obtained by assuming that half the emitting volume was filled by the GOES and the RHESSI plasma, respectively (these combined values are shown in Fig.~\ref{fig:eth_goes}). 

It is well known that RHESSI always yields higher temperatures and lower emission measures than GOES \citep[cf.][]{Battaglia05,Ryan14,Warmuth16a}. These differences arise because flares are not truly isothermal and the RHESSI temperature response is weighted toward higher temperatures than GOES. The net effect on thermal energies is shown in Fig.~\ref{fig:eth_hxsx}(a), where we show the GOES-derived peak thermal energies normalized by those derived based on RHESSI as a function of GOES class for the events of WM16. On average, the GOES-derived energies are higher by a factor of 1.4. This ratio does not depend on flare class. The corresponding ratio of the combined thermal energy, shown in Fig.~\ref{fig:eth_hxsx}(b), shows the corresponding ratios for the combined thermal energy, which is higher on average than the RHESSI-derived energy by a factor of 1.7.

IC+14 went beyond the isothermal approximation and computed thermal energies from a DEM constrained both from EUV imaging with AIA and RHESSI HXR spectra. In principle, this should provide more realistic estimates of the thermal energy. The authors also computed the thermal energies corresponding to an isothermal fit, and the multithermal to isothermal energy ratio is shown in Fig.~\ref{fig:eth_hxsx}(c). The energies are comparable, which can be explained by the fact that most of the thermal energy is contributed by hot plasma that is constrained by RHESSI data (cf. the discussion in IC14). The different data source and thermal model employed by IC14 therefore cannot explain the excess over extrapolations of the energies of E+12 and WM16.

Finally, A+17 derived the multithermal energy at flare peak time using a spatial synthesis method that fits a Gaussian DEM to each spatial pixel of a set of AIA images in the six coronal wavelengths. Figure~\ref{fig:eth_goes} demonstrates that the resulting energies are about an order of magnitude higher than those derived from RHESSI and GOES. A potential explanation for this could be a severe underestimation of thermal energies due to the isothermal assumption of E+12 and WM16. However, IC14 showed that isothermal and multithermal energies can be quite comparable. While this was compared only for microflares, larger flares tend to be hotter, so that the thermal energy will be even more dominated by material that is seen in X-rays, and thus no increased mismatch is to be expected.

An alternative explanation is that the method of A+17 overestimated the thermal energies. IC14 showed that a single-Gaussian DEM profile cannot decrease sufficiently steeply at high temperatures in order to be compatible with RHESSI X-ray observations. While the multiple Gaussians used by A+17 will mitigate this issue to some extent, we note that in fact most DEM reconstruction methods tend to derive too much plasma at high temperatures, resulting in excessive X-ray emission \citep[cf.][]{Su18}. This notion is supported by the fact that many thermal energies of A+17 are on the same order as the bolometric energy, or even exceed it.

\subsubsection{Source volumes}
\label{sec:vol}

The volume that is required for the computation of the thermal energy is usually decomposed into an apparent volume $V$ derived from EUV or X-ray imaging of the thermal source and a volume filling factor $f$ that accounts for the possibility that the emitting plasma may only occupy a fraction of the apparent volume.

\begin{figure}
\resizebox{\hsize}{!}{\includegraphics{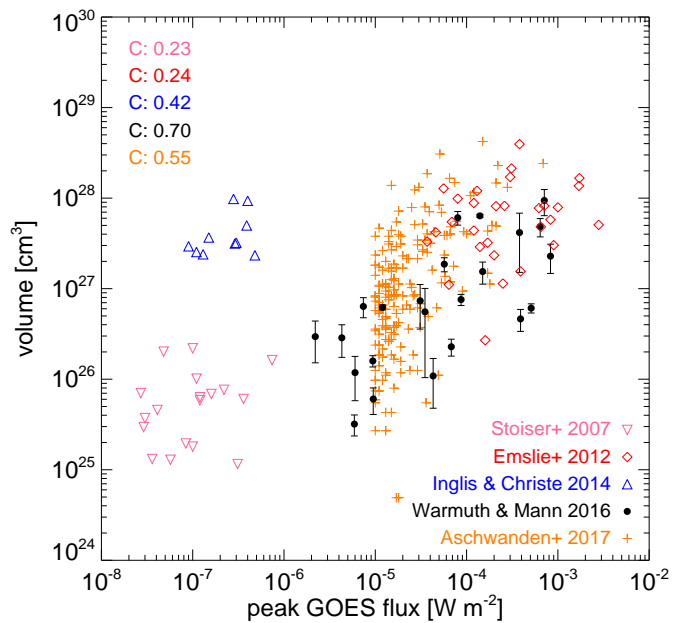}}
\caption{Thermal source volume plotted vs. peak GOES flux for all five studies.}
\label{fig:vol_goes}
\end{figure}

With the exception of S+07 and A+17, all studies used the area of the thermal X-ray source as imaged by RHESSI to constrain the apparent volume, and adopted $f=1$. S+07 derived the volume by assuming a semicircular loop with a cross-sectional area and loop length given by areas and separations of the footpoint brightenings observed at 1600~\AA\ by TRACE, while A+17 estimated the volume from the flare area that was measured above some appropriate threshold in the emission measure per (macro)pixel that results from the spatial synthesis DEM method. 

In Fig.~\ref{fig:vol_goes} we plot the volumes used to derive the peak thermal energies as a function of GOES peak flux. All volumes show considerable scatter, for example, flare volumes span a range of one order of magnitude for the same GOES class for most of the studies, and even two orders of magnitude for A+17. Despite this scatter, all studies show at least a moderate correlation between source volume and GOES class. 

It is noteworthy that the volumes of E+12 and WM16 on the one hand and A+17 on the other hand are generally consistent, although they were derived with two completely different methods. Conversely, the microflare volumes of IC14 are one to two orders of magnitude larger than would be expected from the other four studies. Interestingly, this corresponds to the amount that is required to explain the elevated thermal energies of IC14 discussed in Sect.~\ref{sec:th_scaling}. While these sources could in principle be intrinsically large, their disagreement with the other studies suggests that an overestimation of source size is more likely. This may have been caused by an insufficient accounting for the tendency of the CLEAN imaging algorithm to provide systematically larger source sizes \citep[e.g.][]{Warmuth13a}. Another possibility might be that the difficulty RHESSI has in resolving small sources, which has been demonstrated for nonthermal sources \citep[cf.][]{Dennis09,Warmuth13b}, might also apply to the potentially small thermal sources in microflares.

\subsubsection{Filling factors}
\label{sec:filling}

While all studies assumed a filling factor of unity, we nevertheless have to consider the validity of this assumption, and the consequences that result if this does not hold. A filling factor below unity can affect energy partition because it decreases the thermal energies by a factor of $f^{1/2}$, while it increases the radiative losses by $f^{-1/2}$. In the literature, very diverse results are reported on $f$. While studies in the EUV have tended to yield very low values, that is, $0.001 < f < 0.1$ \citep[e.g.][]{Aschwanden08b}, X-ray observations have generally given higher values in the range of $0.1 < f < 1$ \citep[e.g.][]{Jakimiec11,Guo12}. 

As pointed out by \citet{Caspi14a} and WM16, $f$ can also be constrained by the requirement that the flare plasma has to be magnetically contained, that is, the plasma beta has to be smaller than unity. Thus the required coronal magnetic field strength is dependent on the filling factor according to $B_\mathrm{cor} \sim f^{1/4}$. Measuring magnetic field strengths in solar flares is challenging, but different techniques consistently demonstrate that $B_\mathrm{cor}$ is on the order of a few 100~G in strong flares. For example, in the X8.2 flare of \mbox{2017 September 10} field strengths of 520~G and 148~G at heights (above the limb) of $\approx$20 and 30~Mm were derived by \citet{Gary18} from observation of gyrosynchrotron emission. In the same event, \citet{Kuridze19} obtained 420~G and 350~G at heights of 15 and 25~Mm from spectropolarimetry. We can now compare the required field strengths to these rather firm constraints. Already for $f=1$, the flares of \citet{Caspi14a} require $B_\mathrm{cor}$ of up to 160~G, while WM16 derive values of up to 370~G. This would rule out filling factors below 0.1. This result is supported by spectroscopic observations using density-sensitive lines \citep[e.g.][]{Milligan12}.
We conclude that while the filling factor $f$ remains a poorly constrained parameter, it is unlikely that it affects the results on energy partition in a substantial way. 

\subsubsection{Radiative energy loss}
\label{sec:radloss}

The peak thermal energy is only a lower limit of the energy that is required to generate and sustain the thermal plasma due to energy-loss processes. Deriving the true thermal-nonthermal energy partition thus requires quantifying these losses and assessing their importance. A convenient measure for this is the ratio between the energy loss (integrated over the event) and the peak thermal energy, and we focus on this property in the following discussions.

\begin{figure}
\resizebox{\hsize}{!}{\includegraphics{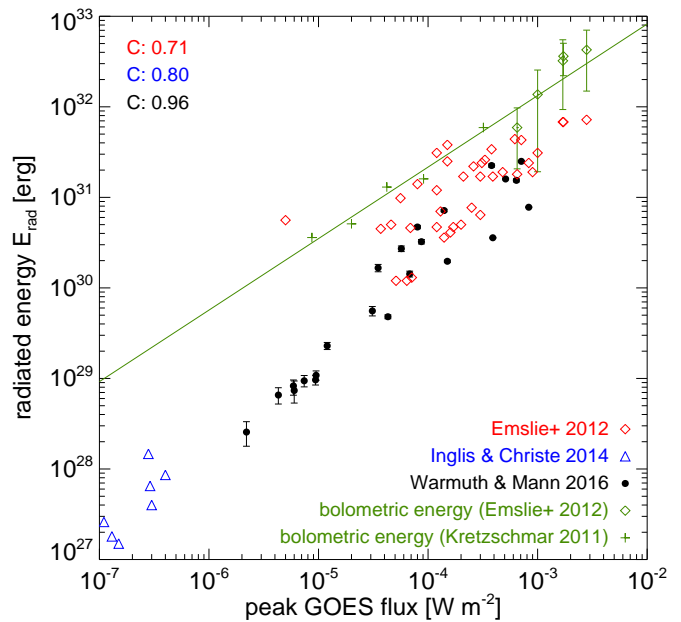}}
\caption{As in Fig.~\ref{fig:eth_goes}, but showing the energy radiated by the hot plasma, $E_\mathrm{rad}$, as a function of peak GOES flux for three studies.}
\label{fig:erad_goes}
\end{figure}

The radiative losses of the hot plasma, $E_\mathrm{rad}$, have been considered in all studies, with the exception of S+07. Figure~\ref{fig:erad_goes} shows $E_\mathrm{rad}$ as a function of GOES peak flux for the different studies. In all cases, $E_\mathrm{rad}$ correlates well with GOES flux. Generally, $E_\mathrm{rad}$ is clearly below the bolometric energy. This is consistent with the understanding that the bolometric emission is dominated by near-UV, white-light, and near-IR emission that originates from comparatively cool material located at lower heights in the solar atmosphere \citep{Woods06,Kretzschmar11}.

\begin{figure}
\resizebox{\hsize}{!}{\includegraphics{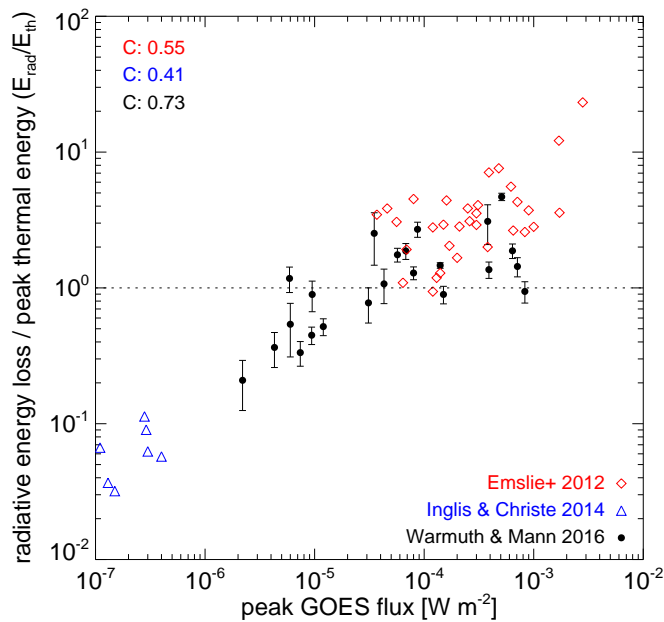}}
\caption{Energy radiated by the hot plasma normalized by the peak thermal energy, $E_\mathrm{rad}$/$E_\mathrm{th}$, vs. peak GOES flux for three studies. The dotted line denotes energy equipartition.}
\label{fig:eradnorm_goes}
\end{figure}

Figure~\ref{fig:eradnorm_goes} shows $E_\mathrm{rad}/E_\mathrm{th}$ for the different studies. In A+17, the radiative losses were not listed numerically, but were shown in a plot, and a (logarithmic) mean ratio of $E_\mathrm{rad}/E_\mathrm{th} = 0.07 \pm 0.06$ was given. The logarithmic mean ratios as well as the corresponding statistical errors of the mean used throughout this work were computed according to the definition in A+17. The corresponding mean ratios for the other studies and methods are listed in Table~\ref{tab:radloss}. Studies using multithermal DEM reconstructions (IC14, A+17) obtain significantly lower ratios ($E_\mathrm{rad}/E_\mathrm{th} <0.1$) than those relying on isothermal fits of X-ray emission (E+12, WM16).

Figure~\ref{fig:erad_goes} shows a distinct trend with GOES class: $E_\mathrm{rad}/E_\mathrm{th}$ increases for higher flare importance. While the radiative losses can be neglected for B-class flares (this even holds when we assume that the thermal energies of IC14 are overestimated; cf. the discussion in Sect.~\ref{sec:vol}), they generally dominate the peak thermal energy for X-class flares.

\begin{table}
\caption{Relation between radiative energy loss $E_\mathrm{rad}$ and peak thermal energy $E_\mathrm{th}$ as derived by different studies and methods. Shown are the logarithmic mean ratios, $E_\mathrm{rad}$/$E_\mathrm{th}$, and the correlation coefficient of the logarithms of the two quantities, $C$.}
\label{tab:radloss}
         \centering
\begin{tabular}{lccc}
        \hline
        \hline
            \noalign{\smallskip}              
    study & method & $E_\mathrm{rad}/E_\mathrm{th}$ & $C$\\
    \hline
    E+12 & RHESSI, isothermal & $3.22 \pm 0.12$ & 0.71\\
    IC14 & RHESSI+AIA, isothermal & $0.06 \pm 0.17$ & 0.88\\
    IC14 & RHESSI+AIA, multithermal & $0.06 \pm 0.16$ & 0.89\\
    WM16 & RHESSI, isothermal & $0.27 \pm 0.19$ & 0.94\\
    WM16 & GOES, isothermal & $1.09 \pm 0.17$ & 0.96\\
    WM16 & RHESSI+GOES, bithermal & $1.07 \pm 0.18$ & 0.96\\
    A+17 & AIA, multithermal & $0.07 \pm 0.06$ & n/a\\
    \hline
\end{tabular}
\end{table}

This result can be understood as follows. A flare is heated impulsively and then cools down. In this case, we would expect $E_\mathrm{rad} = E_\mathrm{th}$ (neglecting conductive losses). However, when heating is more gradual, $E_\mathrm{rad} > E_\mathrm{th}$ is observed because more energy is lost before the thermal peak is reached. This is consistent with the results shown above: larger flares usually have longer durations and more extended impulsive phases as well,  therefore $E_\mathrm{rad}$ dominates. The question now is how we can understand $E_\mathrm{rad} < E_\mathrm{th}$ in weak flares. IC14 noted this point and proposed that in addition to strong conductive losses (see Sect.~\ref{sec:condloss}), a filling factor of about $f \approx 5 \times10^{-3}$ would give an equipartition of peak thermal and radiated energy. As we showed in Sect.~\ref{sec:filling}, the latter explanation may be inconsistent with the magnetic field strength required to contain the plasma. A final possibility is that the thermal energy is overstimated (cf. Sect.~\ref{sec:vol}).

Finally, we address the low value of $E_\mathrm{rad}/E_\mathrm{th}$ derived by A+17. This could have resulted from the thermal energies that are systematically higher than those of the other studies by about an order of magnitude. Moreover, A+17 found that $E_\mathrm{rad}/E_\mathrm{th}$ is anticorrelated with $E_\mathrm{th}$, which is in contrast to the trend seen in the other studies. A+17 argued that this is qualitatively consistent with cooling models that predict that radiative and conductive losses are anticorrelated at higher temperatures \citep[e.g.][]{Cargill95}. However, the radiative loss rate is not strongly dependent on temperature, and the higher emission measure in large flares usually overcompensates for this. It appears that A+17 derived systematically lower radiative losses than E+12 and WM16 for large events: the maximum in A+17 is $E_\mathrm{rad} \approx 10^{31}$~erg, while it approaches $10^{32}$~erg in E+12. The cause for this is unclear because both E+12 and A+17 used the same method to derive $E_\mathrm{rad}$. 

\subsubsection{Conductive energy loss}
\label{sec:condloss}

\begin{figure}
\resizebox{\hsize}{!}{\includegraphics{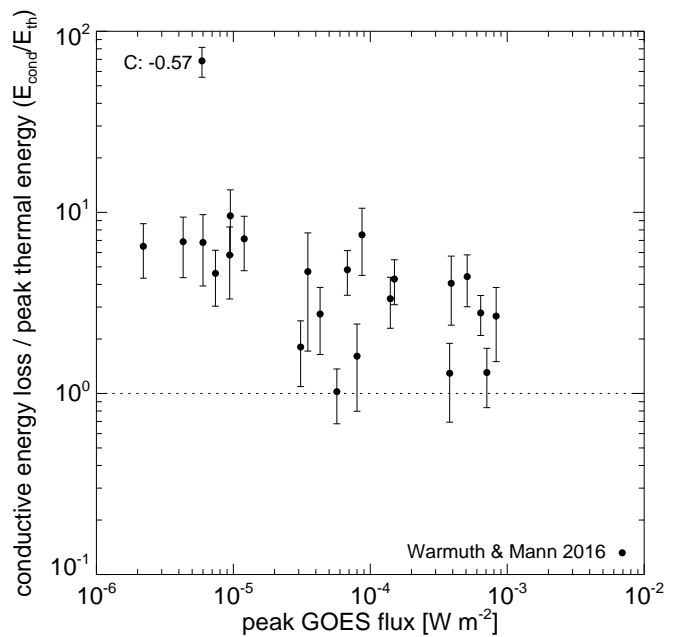}}
\caption{As in Fig.~\ref{fig:eradnorm_goes}, but showing conductive energy loss of the hot flare plasma normalized by the peak thermal energy, $E_\mathrm{cond}$/$E_\mathrm{th}$, as a function of peak  GOES flux as derived by WM16.}
\label{fig:econd_goes}
\end{figure}

While conductive losses have been considered by many authors to investigate flare thermal evolution \citep[e.g.][]{Aschwanden01,Ryan13} and by several case studies of chromospheric evaporation \citep[e.g.][]{Battaglia09}, WM16 provided the first systematic treatment of conductive losses in terms of energetics. Conversely, conduction has been neglected by the other three studies of energetics discussed here.

Applying Spitzer heat conduction \citep{Spitzer62}, WM16 found very large conductive losses, and the corresponding logarithmic mean ratios of $E_\mathrm{cond}$/$E_\mathrm{th}$ are listed in Table~\ref{tab:condloss} for the three different methods used by WM16. In particular, the relative importance of the conductive losses is dependent on GOES class, which is illustrated in Figure~\ref{fig:econd_goes} (the energies plotted refer to the combination of RHESSI- and GOES-derived plasma parameters). For C-class flares, the conductive losses exceed the peak thermal energies by up to one order of magnitude, and this ratio decreases for larger flares. This is the opposite of what was found for the radiative losses (cf. Fig.~\ref{fig:eradnorm_goes}).

\begin{table}
\caption{As in Table~\ref{tab:radloss}, but showing the relation between conductive energy loss $E_\mathrm{cond}$ and peak thermal energy $E_\mathrm{th}$ as derived by WM16 using three different methods for determining the thermal plasma parameters.}
\label{tab:condloss}
         \centering
\begin{tabular}{lccc}
        \hline
        \hline
            \noalign{\smallskip}              
    study & method & $E_\mathrm{cond}/E_\mathrm{th}$ & $C$ \\
    \hline
    WM16 & RHESSI & $9.29 \pm 0.20$ & 0.85\\
    WM16 & GOES & $2.63 \pm 0.21$ & 0.83\\
    WM16 & RHESSI+GOES & $4.16 \pm 0.20$ & 0.85\\  
    \hline
\end{tabular}
\end{table}

The large conductive losses were the main reason for the conclusion in WM16 that the nonthermal energy is only sufficient to power the thermal flare component in the larger events. We need to determine whether such large losses are realistic.

The estimation of the conductive loss involves several uncertainties. To arrive at the conductive energy loss rate, the conductive flux density has to be integrated over the cross-sectional area of the coronal loop footpoints. WM16 took this as the HXR footpoint area. However, RHESSI may have issues with properly measuring footpoint sizes \citep{Dennis09,Warmuth13b}, which could lead to an overestimate of the conductive losses. Conversely, the HXR footpoints map the area where nonthermal electrons are currently precipitating, which may only comprise a fraction of a flaring loop filled with hot plasma. If this effect dominates, then conductive losses would be underestimated.

Another source of uncertainty is the thermal gradient length, which was taken as the half-loop length by WM16. It can be argued that significantly shorter lengths \citep[i.e. corresponding to the transition region height; cf.][]{Fletcher13} could be appropriate, which again would result in even larger losses.

A more fundamental issue is the validity of using Spitzer heat conductivity. Under typical solar flare conditions, the heat flux is usually saturated \citep[e.g.][]{Campbell84,Karpen87}, and this limitation has been accounted for by WM16. However, an additional effect was not considered by WM16 that has been pointed out by \citet{Brown79}: turbulence generated in a flaring loop can scatter electrons. Several recent studies have demonstrated that this can drastically suppress the parallel heat flux \citep[cf.][]{Bian16a,Emslie18,Roberg18}. A high level of turbulence could thus effectively switch off conduction as a loss term. As a consequence, the heating requirements for the thermal flare component in WM16 would drop, in particular for the smaller flares, for which conduction was found to be more important. This means that one of the main conclusions of WM16 would have to be modified: if conduction is suppressed, the nonthermal energy input can account for the thermal requirements even in smaller flares.

While there are arguments for a strong suppression of the conductive heat flux, there is also observational evidence for conduction-driven evaporation \citep[e.g.][]{Antiochos78,Zarro88,Czaykowska01,Battaglia09,Battaglia15}, which suggests that the suppression may not be as severe as predicted, or at least not in all cases. This opens yet another question about the role of conduction in energy partition. In assuming that the energy transported to the deeper atmospheric layers is completely radiated away, WM16 have treated conduction solely as a loss term. However, as the evidence for conduction-driven evaporation shows, at least part of the energy conducted to the chromosphere might be spent in generating newly heated and subsequently evaporated plasma. This would amount to a ``reprocessing'' of the conductive losses, and would in effect lower the total heating requirements of the hot plasma.
We conclude that while conductive losses may potentially be very important for the energy partition and transport, several open questions remain that will have to be properly addressed before any reliable quantitative assessment can be made.

\subsection{Nonthermal issues}
\label{sec:nth}

\subsubsection{Scaling of energy in nonthermal electrons with GOES class}
\label{sec:nth_scaling}

\begin{figure}
\resizebox{\hsize}{!}{\includegraphics{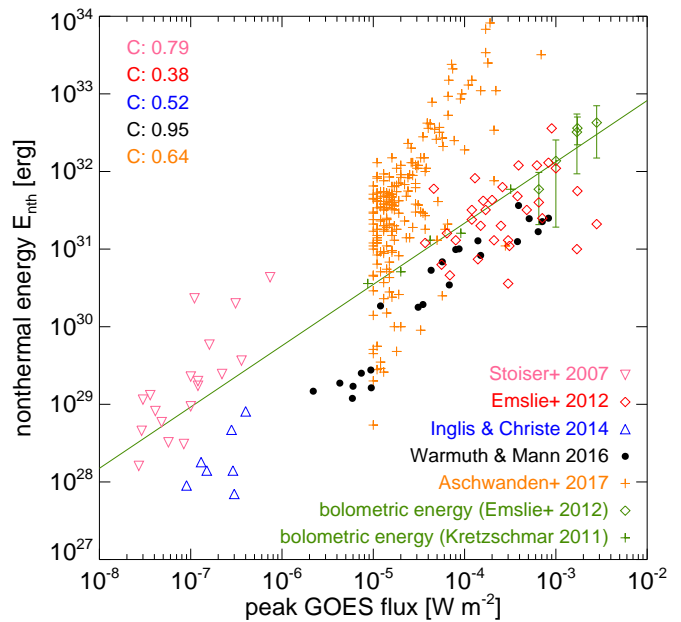}}
\caption{As in Fig.~\ref{fig:eth_goes}, but showing the energy in nonthermal electrons, $E_\mathrm{nth}$, plotted vs. peak GOES flux for all five studies.}
\label{fig:enth_goes}
\end{figure}

\begin{figure*}
\centering
\includegraphics[width=14cm]{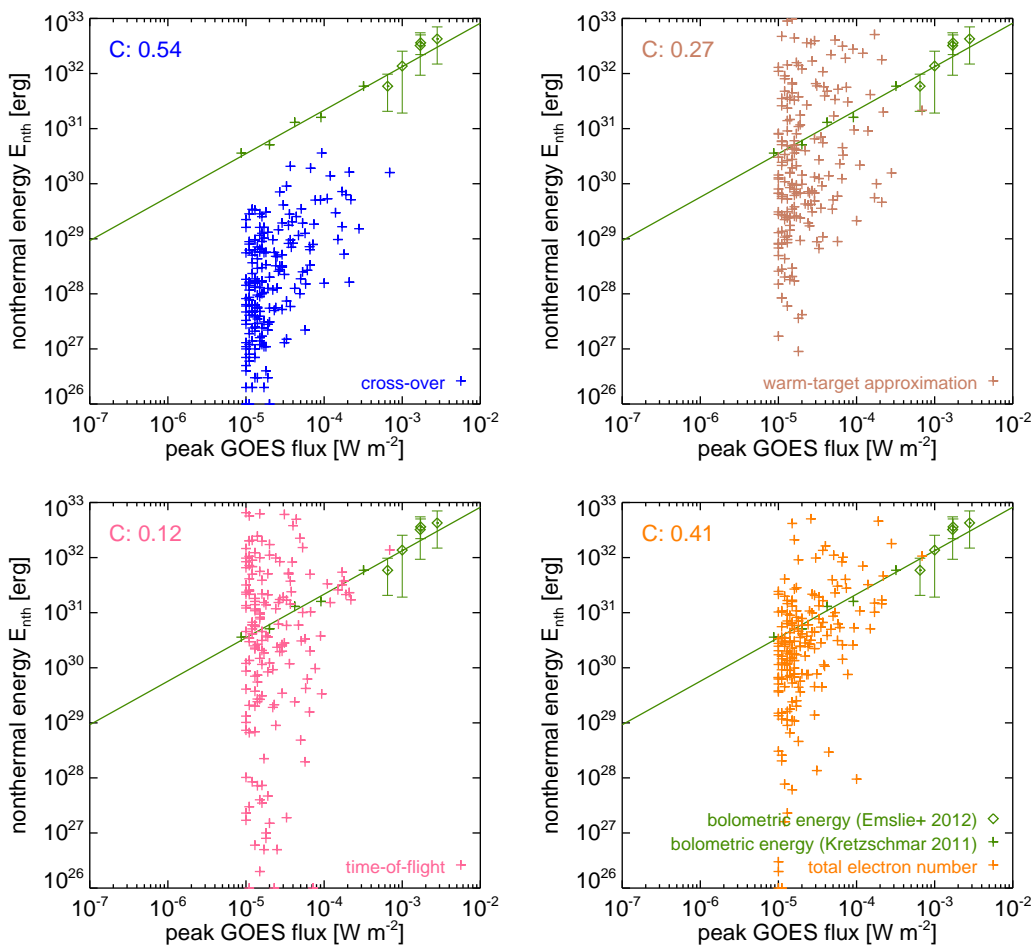}
\caption{As in Fig.~\ref{fig:enth_goes}, but showing the energy in nonthermal electrons, $E_\mathrm{nth}$, as a function of peak GOES flux as derived with four different methods used by \citet{Aschwanden19}.}
\label{fig:enth_goes_a19}
\end{figure*}

The nonthermal component was derived from a much more homogeneous data set than the thermal component in the observations: all studies exclusively used HXR spectroscopic data obtained with RHESSI to constrain the energy in accelerated electrons. All studies used the collisional thick-target model to derive the spectrum of the injected electrons. While E+12, IC14, and WM16 followed the standard practice of using the highest low-energy cutoff consistent with the data, A+17 constrained the cutoff with an analytical approximation of the warm-target model developed by \citet{Kontar15}. S+07 adopted a cutoff of 10~keV for all events, a value which was found to be consistent with the HXR spectra.

In Figure~\ref{fig:enth_goes} we plot the energy in nonthermal electrons, $E_\mathrm{nth}$, derived by the five studies as a function of GOES peak flux. All studies show a correlation with GOES class, but at a slightly lower level than the thermal energies (cf. Fig.~\ref{fig:eth_goes}). The energies of E+12 and WM16 are consistent, and the energies of IC14 also agree with extrapolations of E+12 and WM16 to lower GOES classes. The values derived by these three studies are consistent with the bolometric energy, and there is a trend for the nonthermal energy to decrease with respect to the bolometric energy in smaller flares (IC14 and WM16). In contrast, the nonthermal energies of S+07 and A+17 are significantly higher. Particularly in the latter case, $E_\mathrm{nth}$ is more than an order of magnitude higher than $E_\mathrm{bol}$ in the majority of the events. This strongly suggests that the nonthermal energies have been overestimated by A+17 (and to a lesser degree by S+07) because it is impossible that such a huge amount of energy is injected into the solar atmosphere without being thermalized and subsequently radiated away, thus being detected in the bolometric emission. This clearly demonstrates the benefit of using the bolometric energy as an independent constraint on energetics.

Following up on A+17 and using the same event sample, A+19 have determined the low-energy cutoff and the nonthermal energy using four different methods, which are discussed in Sect.~\ref{sec:elc}. The relation of the corresponding nonthermal energies with GOES class is plotted in Fig.~\ref{fig:enth_goes_a19}. With the exception of the spectral cross-over method (also applied by E+12, IC14, and WM16), all methods yield a very low correlation of the injected energy with GOES peak flux and generally show extremely strong scatter (up to seven orders of magnitude) for events of comparable importance. On average, the derived energies are consistent with (i.e., lower than) the bolometric energy, which contradicts the result of A+17. Moreover, the different methods yield quite consistent nonthermal energies, with the exception of the cross-over method, which gives substantially lower energies.

\subsubsection{Spectral model}
\label{sec:nth_model}

All five studies considered here have used the collisional thick-target bremsstrahlung model \citep{Brown71} to derive the flux (and hence the kinetic power) of the injected electrons from observed HXR photon spectra. The main difference was the way in which the low-energy cutoff of the nonthermal electron distribution was constrained, which is addressed in the following Sect.~\ref{sec:elc}. S+07 fit a photon power law to the spectra that was then converted into a thick-target electron flux.

In the classical thick-target model, the background plasma is considered to be ``cold'' in the sense that the thermal speed of the particles is much slower than the speed of the injected electrons. However, this assumption becomes invalid when the target is heated during the flaring process. Thus, \citet{Kontar15} have recently developed the warm-target model, and extension of the cold-target model that takes into account the physics of collisional energy diffusion and thermalization of fast electrons in the background plasma. As additional input, the warm-target model requires the target temperature, density, and length, which means that imaging observations are required in addition HXR spectra.

The warm-target model can be employed to obtain an upper estimate for the injected nonthermal energy because in contrast to the cold-target model, the low-energy cutoff cannot be made arbitrarily small: the thermalized electrons leave their signature on the photon spectrum. This approach was taken by \citet{Kontar19} in a detailed study of the peak of a single flare, but has not been applied to a larger event sample so far.

\subsubsection{Low-energy cutoff}
\label{sec:elc}

The low-energy cutoff $E_\mathrm{C}$ is the crucial parameter for constraining the energy content of the injected electrons. It is also the most elusive parameter because its spectral signature (a flattening of the photon spectrum below the break) is usually masked by the much stronger thermal emission. There are a few exceptions to this: in early impulsive events \citep{Sui07} and in flares with late impulsive peaks \citep{Warmuth09b,Ireland13}, the low-energy cutoff can be observed directly in the spectra. The standard practice to cope with the masking of the cutoff is to use the highest low-energy cutoff that is still consistent with the data, and thus obtain a lower estimate for the flux and power of the energetic electrons \citep[see e.g.][]{Holman11}. This was done by E+12, IC14, and WM16. In addition, it was one of the methods used by A+19, where it is referred to as the spectral cross-over method. S+07 used a modification of this approach: they obtained the spectral cross-over in the range of 9--12~keV, but adopted a fixed cutoff of 10~keV for all their events.

\begin{figure}
\resizebox{\hsize}{!}{\includegraphics{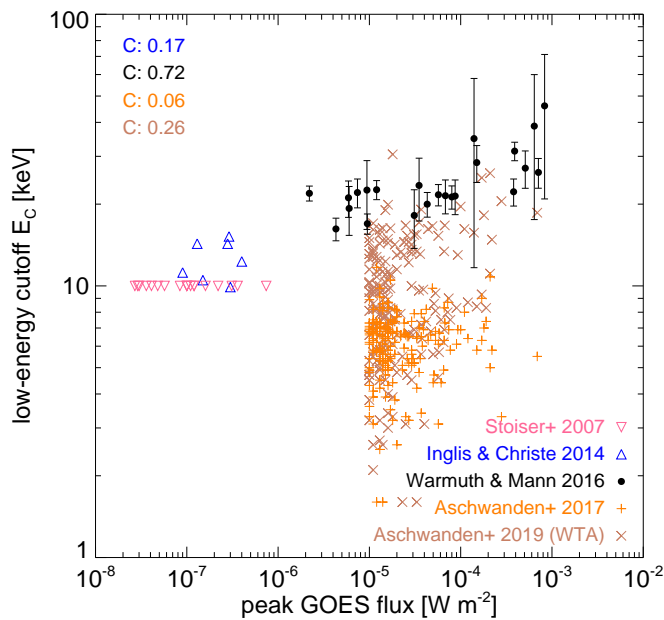}}
\caption{Low-energy cutoffs $E_\mathrm{C}$ plotted vs. GOES peak flux as derived by with the spectral cross-over method by S+07, IC14, and WM16 (here, the mean and standard deviation of the cutoff is shown for each flare), and with the warm-target approximation by A+17 and A+19.}
\label{fig:elc_goes}
\end{figure}

The lower limit on nonthermal energy provided by the cross-over method is to some degree dependent on the way in which the thermal part of the X-ray spectrum is fit. While all studies used an isothermal model, IC14 additionally considered the multithermal case. This resulted in lower nonthermal energies, with a logarithmic mean of $E_\mathrm{nth,multi}/E_\mathrm{nth,iso} = 0.49 \pm 0.34$. A multithermal fit thus gives the absolute lower limit on the electron energy content. Although the median difference of the cutoff energies in IC14 was only 2.5~keV, this nevertheless resulted in a substantial effect on the total energies. This is generally the case for the steep spectral indices that are characteristic for microflares \citep[i.e. electron spectral indices $\approx 8$; cf.][]{Christe08}. This strong sensitivity might be one reasons for the apparent overestimation of nonthermal energies by S+07, where the universally adopted cutoff of 10~keV is a few keV lower than in most of the events of IC14 (cf. Fig.~\ref{fig:elc_goes}).

A+17 applied a different method to constrain $E_\mathrm{C}$. They used an approximation from the warm-target model of \citet{Kontar15} that provides an effective low-energy cutoff that is linearly dependent on the electron spectral index and the warm-target temperature. As temperature, 8.6~MK was adopted for all flares and time intervals, which corresponded to the mean EM-weighted temperature derived from the DEM reconstructions. This approach has resulted in significantly lower cutoff energies (with a mean of $E_\mathrm{C} = 6.2$~keV) than in studies using the cross-over method (with typically $E_\mathrm{C} > 10$~keV), and therefore produced higher nonthermal energies. To illustrate this, we show the low-energy cutoffs determined by S+07, IC14, WM16, and A+17 as a function of GOES peak flux in Fig.~\ref{fig:elc_goes}.

A+19 expanded on the issue of the low-energy cutoff by applying four different constraints on $E_\mathrm{C}$ to the event sample of A+17. These were (a) the familiar cross-over method (CO), (b) the warm-target approximation used by A+17 (WTA), but now using the geometric mean of RHESSI- and AIA-derived temperatures measured in each individual event, (c) a model based on the equivalence of the time-of-flight (TOF) of electrons and the collisional deflection time, and (d) the total electron number model (TEN), which relies on the number of electrons that is available for acceleration in the  flaring region.

A+19 found that generally the WTA, TOF, and TEN models yield consistent results on the cutoff energies, which are found to be around 10~keV on average. For comparison, we have included in Fig.~\ref{fig:elc_goes} the WTA values of A+19. The effect of the higher temperatures used by A+19 is evident when it is compared to the results of A+17. While these cutoffs are significantly higher than the average value of A+17, they correspond to the lowest values for $E_\mathrm{C}$ derived by IC14 and WM16 from the cross-over method. 

We note that the nonthermal energies derived by A+17 and A+19 generally show a significantly stronger scatter for flares of comparable importance than the other studies. A possible cause for this might be the automatic fitting procedures that had to applied because of the large sample size.

\subsubsection{Time integration of nonthermal input}
\label{sec:time_nth}

While a single HXR spectrum is sufficient to derive a thermal energy, a thick-target fit only yields the power of the injected electrons. Obtaining the nonthermal energy thus requires a time integration over the event (similar to the radiative and conductive losses), or at least an assumption about the duration of the nonthermal energy input. E+12, WM16 and A+17 have all split each individual flare into time bins (typically of 20~s duration), performed spectral fits for each bin, and integrated over the whole event. 

In contrast, low count rates and short durations of the microflares studied by S+07 and IC14 demanded a different approach, and in both cases only a single spectrum was fit for each flare. S+07 fit a spectrum with 12~s integration time obtained around the HXR peak and integrated the nonthermal power over the time of detectable emission above 10~keV under the assumption of a triangular time profile. IC14 used spectra with 60~s integration time and integrated the nonthermal power over the same duration. When a single spectral fit is used, it results in a less realistic characterization of the nonthermal emission. This may have contributed to the large discrepancy between the nonthermal energies derived by S+07 and IC14.

\subsubsection{Contribution of ions}
\label{sec:ions}

When we discussed nonthermal energetics, we only considered the energetics of electrons so far. To quantify the thermal-nonthermal energy partition and the acceleration efficiency, we need to consider the contribution of energetic ions as well. Unfortunately, the energy content of ions is even less well constrained than that of electrons. In strong flares, observation of gamma-ray lines allows the determination of the energetics of $>$1~MeV ions \citep[e.g.][]{Lin03,Emslie04}. Based on fluences in the neutron-capture line \citep{Shih09}, E+12 concluded that the energy content of electron and ions is generally comparable within an order of magnitude (with a logarithmic mean of $E_\mathrm{nth,i}/E_\mathrm{nth,e} = 0.34 \pm 0.50$). The partition does not show a dependence on flare importance.

A pragmatic approach for determining the total nonthermal energy would thus be to simply multiply the energy in electrons by a factor of 2, while having in mind that the contribution by ions is not well determined. In particular, the ion spectrum below 1~MeV is not constrained at all, therefore the total energy in nonthermal particles might well be higher. 

\subsection{Additional flare energetics not considered so far}
\label{sec:addit}

Our discussion of flare energetics has focused on thermal energies and energy losses of the hot plasma, and on the energy in nonthermal particles. However, several additional energy components may be important for understanding energy partition in general. We briefly address them.

The hot thermal plasma not only contains thermal energy, but also \textit{\textup{gravitational energy}} (because the bulk of the flare plasma has to be transported from the chromosphere to larger coronal heights), and \textit{\textup{kinetic energy}} of the evaporation \textup{\textup{flows}}. WM16 demonstrated that these energies are far lower than the thermal energy. The kinetic flow energy is more than an order of magnitude lower than the peak thermal energy, and the gravitational energy is a full two orders of magnitude lower.

A different type of flow is directly associated with the reconnection process. The reconnection outflow jets contain the kinetic energy of the bulk flow, and additionally, the kinetic energy associated with \textup{}turbulence. Several models assume that the energy required for particle acceleration is either taken from the bulk outflow \citep[through acceleration at shock waves, e.g.][]{Mann09} or from turbulent flows within the outflow \citep[through stochastic acceleration, e.g.][and references therein]{Petrosian12}. In this scenario, the flow energy should not be added to the thermal-nonthermal energy budget. It rather represents the reservoir from which the energy in nonthermal particles is supplied and is thus an intermediate step between the dissipated magnetic energy and the accelerated particles. 

For this model to be feasible, the flow energy should at least be comparable to the total nonthermal energy. This is indeed supported by observations. \citet{Warmuth09a} constrained the density and speed of the outflow with radio observations of termination shocks and found the kinetic energy sufficient to account for the nonthermal particles. Conversely, \citet{Kontar17} derived the kinetic energy associated with small-scale turbulent mass motions from the nonthermal velocity broadening of EUV lines and found it to be very consistent with the energy in accelerated particles.

Another component we did not consider is the warm coronal plasma (i.e., plasma at temperatures of below 5~MK that does not produce significant X-ray emission) that produces a prolonged secondary peak in EUV emission during the gradual phase of a small fraction of flares (13\%), the so-called \textit{\textup{EUV late phase}} \citep[cf.][]{Woods11}. This component, located in distinct loop systems,  implies additional heating requirements. Just considering radiative losses in EUV, it has been shown that the EUV late phase can be up to four times more energetic than the main phase \citep{LiuK15}, and numerical modeling suggests that the peak heating rate for the late-phase loops may be at least as high as for the main flaring loops \citep{DaiY18}.

It has been proposed that the heating required by the EUV late phase may be supplied by the thermalization of the energy contained in a flux rope after a failed eruption \citep{WangY16}. If this is the case, then the existence of the EUV late phase does not directly affect the considerations on flare energy partition we made here because it rather relates to the CME energetics. However, this does demonstrate the close interplay between the processes we tend to divide into flare and CME. A full understanding of solar eruptive events will require taking into account both manifestations of the phenomenon.

\subsection{Thermal-nonthermal energy partition: dependence on flare importance}
\label{sec:th_nth}

After discussing in detail the issues that affect the derivation of thermal and nonthermal energetics in solar flares, we return to the basic question of energy partition. Because all thermal energetics and most of the nonthermal energies scale rather well with GOES peak flux, we now assess the dependence of the energy balance on flare importance.

We first consider the relation between peak thermal energy $E_\mathrm{th}$ and energy in nonthermal electrons,  $E_\mathrm{nth}$. Table~\ref{tab:nth_th} shows the mean (logarithmic) ratios of $E_\mathrm{nth}/E_\mathrm{th}$ as derived by all studies and methods considered here, and the correlation between the two parameters. While S+07, E+12, WM16, and A+17 reported that the energy in electrons is more than sufficient to account for the thermal energy, both IC14 and A+19 reported a deficit of electrons. The correlation between the two quantities is lower than the correlation of each individual energy component with peak GOES flux (cf. Figs.~\ref{fig:eth_goes} and \ref{fig:enth_goes}), which is not surprising considering the uncertainties involved in the derivation of both thermal and nonthermal components.

\begin{table}
\caption{Relation between energy in nonthermal electrons $E_\mathrm{nth}$ and peak thermal energy $E_\mathrm{th}$ as derived by different studies and methods. Shown are the logarithmic mean ratios, $E_\mathrm{nth}$/$E_\mathrm{th}$, and the correlation coefficient of the logarithms of the two quantities, $C$.}
\label{tab:nth_th}
         \centering
\begin{tabular}{lccc}
        \hline
        \hline
            \noalign{\smallskip}                   
    study & method & $E_\mathrm{nth}/E_\mathrm{th}$ & $C$\\
    \hline
S+07 & CO & $28.3 \pm 0.29$ & 0.76 \\
E+12 & CO & $6.78 \pm 0.21$ & 0.32 \\
IC14 & CO, isotherm. & $0.55 \pm 0.24$ & 0.75 \\
IC14 & CO, multitherm. & $0.28 \pm 0.21$ & 0.84 \\
WM16 & CO, RHESSI & $4.59 \pm 0.17$ & 0.93 \\
WM16 & CO, GOES & $3.27 \pm 0.18$ & 0.93 \\
WM16 & CO, combined & $2.71 \pm 0.18$ & 0.93 \\
A+17 & WTA & $6.72 \pm 0.15$ & 0.42\\
A+19 & CO & $0.003 \pm 0.17$ & 0.34\\
A+19 & WTA & $0.42 \pm 0.26$ & 0.05\\
A+19 & TOF & $0.49 \pm 0.32$ & 0.19\\
A+19 & TEN & $0.36 \pm 0.21$ & 0.34\\
    \hline
\end{tabular}
\tablefoot{The methods used to constrain the low-energy cutoff are spectral cross-over (CO), warm-target approximation (WTA), time-of-flight (TOF), and total electron number (TEN).}
\end{table}

\begin{figure}
\resizebox{\hsize}{!}{\includegraphics{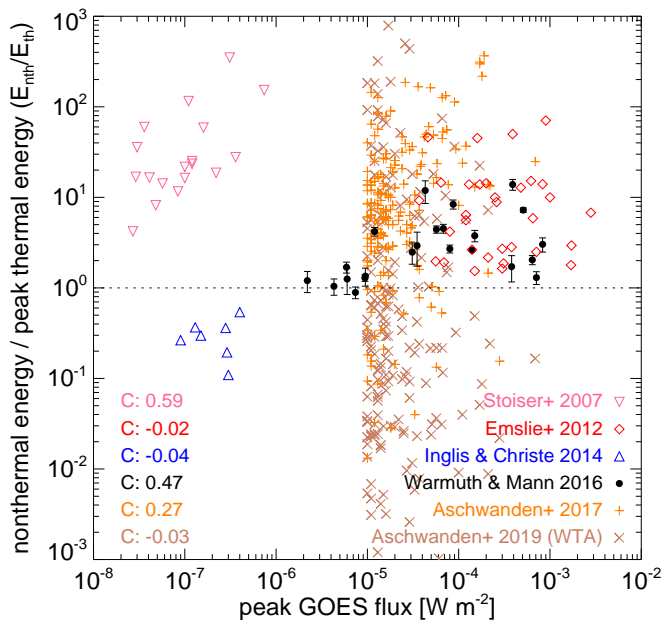}}
\caption{Energy in nonthermal electrons normalized by peak thermal energy, $E_\mathrm{nth}/E_\mathrm{th}$, vs. peak GOES flux as derived by the five studies. In addition, the ratio is shown for the nonthermal energies derived with the warm-target approximation in A+19.}
\label{fig:enth_eth_goes}
\end{figure}

Going beyond mean values, we plot the ratio $E_\mathrm{nth}/E_\mathrm{th}$ as a function of peak GOES flux in Fig.~\ref{fig:enth_eth_goes}. While the individual studies show only low correlations between the energy ratio and GOES class (with the exception of S+07 and WM16), the different results considered together might indicate a trend: the microflares of IC14 all show a deficit of nonthermal energy, the thermal component in larger events of E+12 can all easily be accounted for by the injected electrons, and the flares of WM16 tie these two regimes together. For A+17 and A+19, the scatter is too severe to discern any trend. However, the microflare energetics derived by S+07 are evidently not consistent with this scenario because they suggest a clear dominance of the nonthermal energy.

We thus have two studies of microflares that yield contradicting results on energy partition. As an independent check, we now consider the large statistical study on microflares by \citet{Hannah08}. We did not discussed it here in detail because it lacks the quantification of nonthermal energies (only peak powers of injected electrons were provided). However, we note that the median thermal energy of 9\,161 microflares was 10$^{26}$~erg, while the median peak power in electrons for 4\,236 microflares was 10$^{28}$~erg\,s$^{-1}$. The typical time duration of nonthermal HXR emission in microflares was given as $\approx$10~s, which would correspond to a median electron energy content of 10$^{27}$~erg in the microflare sample. This means that on average, the energetic electron input can account for only $\approx$10\% of the peak thermal energy in microflares. This is consistent with the results of IC14 and extrapolated ratios of WM16. We thus conclude that the nonthermal energies of IC14 are more realistic than those of S+07. This is supported by the disagreement between the nonthermal and the bolometric energies in S+07 (cf. Fig.~\ref{fig:enth_goes}), which suggests an overestimation of $E_\mathrm{nth}$. Still, the discrepancies between the microflare studies clearly illustrate the difficulty in accurately measuring the nonthermal energy in small events, which is caused by poor statistics (low number of nonthermal counts), issues with background subtraction, and the steep spectra that lead to a very high sensitivity of the derived electron flux with respect to the low-energy cutoff.

\begin{figure}
\resizebox{\hsize}{!}{\includegraphics{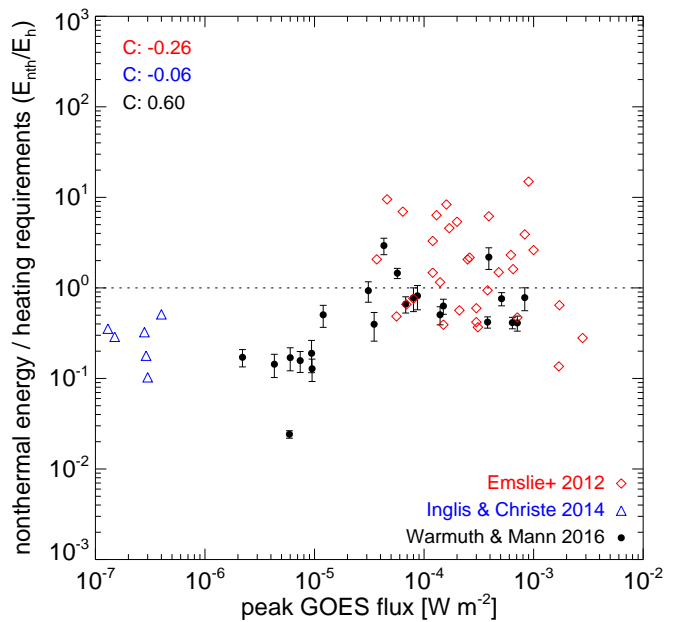}}
\caption{As in Fig.~\ref{fig:enth_eth_goes}, but showing the energy in nonthermal electrons normalized by the total heating requirements, $E_\mathrm{nth}/E_\mathrm{h}$, vs. peak GOES flux as derived by E+12, IC14, and WM16. The energy required to generate and sustain the hot flare component is approximated by adding the peak thermal energy and the radiative losses, $E_\mathrm{th}+E_\mathrm{rad}$, for E+12 and IC14, while it is derived from a time integration of the thermal energy change rate plus the radiative and conductive energy loss rates for WM16.}
\label{fig:enth_eh_goes}
\end{figure}

The overall trend of a decreasing nonthermal-to-thermal ratio with decreasing flare importance is confirmed when the thermal losses are considered as well. In Fig.~\ref{fig:enth_eh_goes} we plot the ratio of nonthermal electron energy over the total heating requirements of the hot plasma (A+17 and A+19 are omitted here because of their large scatter). The latter quantity is approximated by adding the peak thermal energy and the radiative losses, $E_\mathrm{th}+E_\mathrm{rad}$, for E+12 and IC14. For WM16, we used the total heating requirements derived from a time integration of the thermal energy change rate plus the radiative and conductive energy-loss rates.

\subsection{Acceleration efficiency, direct heating, and energy transport}
\label{sec:acceff}

The decreasing nonthermal to thermal energy ratio for weaker flares established in the previous section might be a spurious trend due to two issues connected to the thermal energetics. First, the low nonthermal ratio in IC14 might primarily be due to an overestimate of the thermal energy caused by an oversized source volume (cf. Sect.~\ref{sec:vol}). Second, the decreasing ratio in WM16 is partly due to the conductive losses that are more important for weaker flares. When conduction is suppressed or the losses are reprocessed (cf. Sect.\ref{sec:condloss}), this trend is significantly weaker.

We can avoid these issues when we compare the nonthermal energy to the bolometric radiated energy. We stress once again that $E_\mathrm{bol}$ is a proxy for the total energy that is released in a flare. We thus define $E_\mathrm{nth}/E_\mathrm{bol}$ as the \textit{\textup{nonthermal fraction}}, which is the fraction of energy dissipated that is converted into nonthermal particles (electrons, in our case).

\begin{figure}
\resizebox{\hsize}{!}{\includegraphics{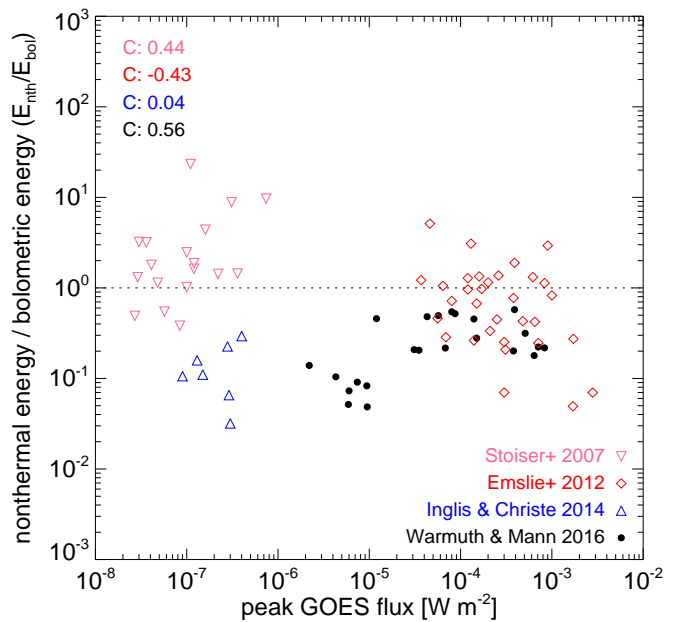}}
\caption{As in Fig.~\ref{fig:enth_eh_goes}, but showing the energy in nonthermal electrons normalized by bolometric radiated energy, $E_\mathrm{nth}/E_\mathrm{bol}$, vs. peak GOES flux as derived by S+07, E+12, IC14, and WM16. $E_\mathrm{bol}$ is derived from a power-law fit to the values obtained by \citet{Kretzschmar11}.}
\label{fig:enth_ebol_goes}
\end{figure}

Table~\ref{tab:nth_bol} shows the logarithmic averages for all studies and methods $E_\mathrm{nth}/E_\mathrm{bol}$ and the correlation of the two quantities, while in Fig.~\ref{fig:enth_ebol_goes} we plot this nonthermal ratio for the individual flares of S+07, E+12, IC14, and WM16 as a function of peak GOES flux (again, the results of A+17 and A+19 are not shown here due to their large scatter). Figure~\ref{fig:enth_ebol_goes} demonstrates that $E_\mathrm{nth}/E_\mathrm{bol}$ also follows the overall trend of a decreasing nonthermal fraction for weaker flares (when S+07 is not considered, where $E_\mathrm{nth}$ has most likely been overestimated): while $E_\mathrm{nth}/E_\mathrm{bol}$ is of the order of unity in X-class flares (but with significant scatter), it is an order of magnitude lower in C-class flares and microflares. This behavior might be an artifact caused by a systematic underestimation of the nonthermal energy in weaker events, as indeed all nonthermal energies plotted in Fig.~\ref{fig:enth_ebol_goes} are lower limits. However, WM16 pointed out that this is  an unlikely scenario because it would require either a systematically lower low-energy cutoff or a higher contribution of ions in weaker flares. For neither scenario do we have observational evidence or a theoretical justification. We therefore identified the changing nonthermal to thermal energy fraction as the main reason for the dissimilar results on energy partition provided by the different studies.

\begin{table}
\caption{As in Table~\ref{tab:nth_th}, but showing the relation between energy in nonthermal electrons $E_\mathrm{nth}$ and bolometric radiated energy $E_\mathrm{bol}$ as derived by different studies and methods. $E_\mathrm{bol}$ is derived from a power-law fit to the values obtained by \citet{Kretzschmar11}.}
\label{tab:nth_bol}
         \centering
\begin{tabular}{lccc}
        \hline
        \hline
            \noalign{\smallskip}                   
    study & method & $E_\mathrm{nth}/E_\mathrm{bol}$ & $C$\\
    \hline
S+07 & CO & $2.05 \pm 0.29$ & 0.44 \\
E+12 & CO & $0.61 \pm 0.21$ & -0.43 \\
IC14 & CO, isotherm. & $0.20 \pm 0.32$ & -0.05 \\
IC14 & CO, multitherm. & $0.11 \pm 0.33$ & 0.04 \\
WM16 & CO & $0.21 \pm 0.17$ & 0.56 \\
A+17 & WTA & $5.42 \pm 0.13$ & 0.39 \\
A+19 & CO & $0.003 \pm 0.16$ & 0.28 \\
A+19 & WTA & $0.34 \pm 0.24$ & 0.06 \\
A+19 & TOF & $0.37 \pm 0.32$ & -0.05 \\
A+19 & TEN & $0.29 \pm 0.21$ & 0.20 \\
    \hline
\end{tabular}
\end{table}

We therefore conclude that in strong flares, the nonthermal energy input is sufficient to generate and sustain the hot thermal plasma, which is consistent with the well-established scenario of chromospheric evaporation driven by electron beams \citep[e.g.][]{Milligan06a,Tian14}. Conversely, in weaker events (microflares and C-class flares), the nonthermal energy input appears to be insufficient. This implies the presence of an additional non-beam heating mechanism, and indeed, several recent studies have found evidence of direct heating of coronal plasma \citep[][WM16, A+17]{Caspi10,Caspi15b}.

We propose that the changing nonthermal fraction might well be due to an acceleration efficiency that increases with flare importance. According to this scenario, the acceleration mechanism works at a low efficiency in weaker flares, which means that a comparatively small fraction of the dissipated magnetic energy is used to accelerate particles, and a larger fraction goes into direct heating. We stress that any acceleration mechanism will always generate an enhanced thermal particle distribution in addition to the nonthermal component. In stronger events, the acceleration process appears to operate at higher efficiency, thus converting a large fraction of the energy released into nonthermal energy.

Several acceleration mechanisms predict varying nonthermal fractions that are consistent with this understanding. For example, the efficiency of shock-drift acceleration of electrons at a termination shock \citep[cf.][]{Mann09,Warmuth09a} is strongly dependent on the temperature and speed of the reconnection outflow jet, which will both be lower in a weaker event. However, heating at the termination shock will be less susceptible to these parameters, and thus a lower nonthermal fraction will result. Another example was presented by \citet{Dahlin16}, who performed kinetic simulations of reconnection and found that the guide field  strongly affects the nonthermal fraction. In the presence of a strong guide field, plasma is predominantly heated, whereas a weak guide field (e.g. in a highly twisted flux rope) allows more efficient acceleration.

The low nonthermal fraction $E_\mathrm{nth}/E_\mathrm{bol}$ in weaker flares has another important consequence. It has been demonstrated that the bolometric emission is dominated by near-UV, white-light, and near-IR radiation, which mainly originates from comparatively cool material located at deeper layers of the solar atmosphere (chromosphere and photosphere). When we accept that the primary energy release takes place in the corona, energy has to be transported to these deeper layers in order to heat the material. Based on energetics, we have shown that electron beams are not sufficient for this in weaker flares. We need an additional energy transport mechanism. Proton beams have been proposed \citep[e.g.][]{Emslie98}, but again it is not evident why smaller events should be proton-dominated (cf. Sect.~\ref{sec:ions}). WM16 have shown that the conductive losses are a viable additional energy transport process that can quantitatively reproduce the bolometric energy. However, this will not work when conduction is suppressed or when a significant fraction of the conductive flux is recycled in the form of conduction-driven evaporation. In this case, the only energy transport mechanism left are magnetohydrodynamic waves \citep[see e.g.][]{Fletcher08,Russell13,Reep16}.


\section{Conclusions}
\label{sec:conc}

We have reviewed in detail five recent studies that determined both the thermal and nonthermal energy content in samples of solar flares: \citet[][S+07]{Stoiser07}, \citet[][E+12]{Emslie12}, \citet[][IC14]{Inglis14}, \citet[][WM16]{Warmuth16b}, and \citet[][A+17]{Aschwanden17a}, with the recent update of \citet[][A+19]{Aschwanden19}. All studies yielded disagreeing results on the energy partition. Our aims were to (a) identify the causes for the disagreements and (b) to investigate whether firmer constraints on energy partition can be achieved by considering the studies together. We summarize our main results as follows.

\begin{enumerate}

\item The largest uncertainty in the determination of the \textit{\textup{thermal energy}} of the hot plasma is the determination of the DEM distribution. According to the method used, the derived energy can be consistent with the value given by an isothermal fit of the X-ray spectrum (IC14) or up to an order of magnitude higher (A+17).
\item Energy losses of the hot plasma are significant. While the studies mostly agree on the \textit{\textup{radiative losses}}, the role of \textit{\textup{conductive losses}} is still a matter of great debate. WM16 have demonstrated that conduction can be a very relevant loss term, but there are arguments that suggest that conduction may be either significantly suppressed or that the losses are recycled through conduction-driven evaporation.
\item The \textit{\textup{nonthermal energy}} in injected electrons is strongly dependent on the poorly constrained low-energy cutoff. The spectral cross-over method gives an lower estimate for the injected energy, and as applied by E+12, IC14, and WM16, this has yielded consistent results. Three alternative methods applied by A+19 have given comparable average energies.
\item The \textup{\textit{\textup{bolometric radiated energy}},} as a proxy for the total energy released in a flare, is a useful independent constraint on both thermal and nonthermal energetics. While the results of most studies are consistent with it, the nonthermal energies A+17 clearly violate this constraint.
\item Generally, thermal and nonthermal energies have shown reasonable correlations with the peak SXR flux as measured by GOES.
\item Considering all studies together, we note that the \textit{\textup{thermal-nonthermal energy partition}} changes with flare importance. In weak flares, there appears to be a deficit of energetic electrons, while the injected nonthermal energy is sufficient to account for the thermal component in strong flares.  This tendency is found for the ratio of nonthermal energy to peak thermal energy, to total heating requirements (including losses), and to the bolometric energy, which to some degree mitigates the uncertainties in the determination of the thermal energetics. Thus the changing energy partition is identified as the main cause of the dissimilar results obtained by the different studies.
\item As a consequence, an additional \textit{\textup{direct heating}} process has to be present, and considering that the bolometric emission originates mainly from deeper atmospheric layers, conduction or waves are required as \textit{\textup{additional energy transport mechanisms}}.
\end{enumerate}

An improvement in our understanding of energy partition in solar flares, and hence flare physics in general, will require several steps. On the thermal side, energetics have to be derived from DEM distributions that are either constrained by both EUV and X-ray data \citep[e.g.][]{Battaglia13,Inglis14,Caspi14b} or reconstructed with algorithms that have been shown to produce results consistent with X-ray spectroscopy \citep[cf.][]{Su18}. Perhaps even more importantly, the importance of conductive losses has to be quantified with more confidence. This will require detailed time-resolved studies of the role of conduction, involving observations, theory, and numerical simulations. 

Improved constraints on the low-energy cutoff will be crucial for the nonthermal component. We propose to conduct systematic studies using the full warm-target model (not just an approximation), which will for the first time provide upper boundaries on the energy in injected electrons (as opposed to the lower boundaries provided  by the common spectral cross-over method). This should be combined with novel approaches, such as using the time profile of X-ray emission at different energies to constrain the cutoff \citep[cf.][]{Dennis19}.
We suggest that all these methods should be applied to flares of various GOES classes in order to ascertain whether a dependence of energy partition on flare importance truly exists.


\begin{acknowledgements}
The work of A. W. was supported by DLR under grant No. 50 QL 1701. We acknowledge support from the International Space Science Institute through the ISSI team
on ``Solar flare acceleration signatures and their connection to solar energetic particles''. We thank Andrew Inglis, Brian Dennis, Gordon Emslie, and Markus Aschwanden for the provision of supplementary data and helpful discussions.
\end{acknowledgements}

\bibliographystyle{aa}
\bibliography{bib_warmuth}

\end{document}